\documentclass[a4paper,11pt]{article}
\usepackage{pos}
\usepackage{sidecap}
\sidecaptionvpos{figure}{m}
\sidecaptionvpos{table}{m}
\usepackage{cleveref}
\usepackage{enumitem}
\setitemize{itemsep=0pt,parsep=1pt,topsep=1pt,itemindent=0pt,leftmargin=9pt}
\usepackage{natbib}
\usepackage{xcolor}

\usepackage{wrapfig}

\usepackage{etoolbox}
\makeatletter
\preto{\@verbatim}{\topsep=0pt \partopsep=0pt }
\makeatother
\AtBeginEnvironment{verbatim}{\setlist[trivlist]{nolistsep}}

\usepackage{listings}

\definecolor{codegreen}{rgb}{0,0.6,0}
\definecolor{codegray}{rgb}{0.5,0.5,0.5}
\definecolor{codepurple}{rgb}{0.58,0,0.82}
\definecolor{backcolour}{rgb}{0.95,0.95,0.92}

\lstdefinestyle{mystyle}{
    backgroundcolor=\color{backcolour},   
    commentstyle=\color{codegreen},
    keywordstyle=\color{magenta},
    numberstyle=\tiny\color{codegray},
    stringstyle=\color{codepurple},
    basicstyle=\ttfamily\tiny,
    breakatwhitespace=false,
    breaklines=true, 
    captionpos=b,
    keepspaces=true,
    numbers=left,
    numbersep=5pt,
    showspaces=false,
    showstringspaces=false,
    showtabs=false,
    tabsize=2
}

\usepackage[font=small,labelfont=bf,tableposition=top]{caption}

\newcommand{\hQpm}{\hat{Q}^{\pm}} 
\newcommand{\hWpm}{\hat{W}^{\pm}}
\newcommand{\hQp}{\hat{Q}^{+}}
\newcommand{\hQm}{\hat{Q}^{-}}
\newcommand{\hQh}{\hat{Q}_h}
\newcommand{\hWp}{\hat{W}^{+}}
\newcommand{\hWm}{\hat{W}^{-}}

\newcommand{\Qpm}{Q^{\pm}}
\newcommand{\Qp}{Q^{+}}
\newcommand{\Qm}{Q^{-}}

\newcommand{\Moe}{M_{oe}}
\newcommand{\Moo}{M_{oo}}
\newcommand{\Meo}{M_{eo}}
\newcommand{\Mee}{M_{ee}}

\title{Twisted mass ensemble generation on GPU machines}

\manuallySeparateAuthors

\author*[a]{Bartosz Kostrzewa}
\author[c]{, Simone Bacchio}
\author[c]{, Jacob Finkenrath}
\author[b]{, Marco Garofalo}
\author[c]{, Ferenc Pittler}
\author[b]{, Simone Romiti}
\author[b]{and Carsten Urbach}
\author{for the ETM Collaboration}

\affiliation[a]{High-Performance Computing and Analytics Lab, Rheinische Friedrich-Wilhelms-Universität Bonn, Friedrich-Hirzebruch-Allee 8, 53115 Bonn, Germany}

\affiliation[b]{Helmholtz-Institut für Strahlen- und Kernphysik (Theorie) and Bethe Center for Theoretical Physics, Rheinische Friedrich-Wilhelms-Universität Bonn, Nussallee 14-16, 53115 Bonn, Germany}
  
\affiliation[c]{The Cyprus Institute, CASTORC, 20 Konstantinou Kavafi Street, 2121 Nicosia, Cyprus}

\emailAdd{bartosz.kostrzewa@uni-bonn.de}
\emailAdd{s.bacchio@cyi.ac.cy}
\emailAdd{j.finkenrath@cyi.ac.cy}
\emailAdd{garofalo@hiskp.uni-bonn.de}
\emailAdd{sromiti@uni-bonn.de}
\emailAdd{urbach@hiskp.uni-bonn.de}

\abstract{
We present how we ported the Hybrid Monte Carlo implementation in the tmLQCD software suite to GPUs through offloading its most expensive parts to the QUDA library.
We discuss our motivations and some of the technical challenges that we encountered as we added the required functionality to both tmLQCD and QUDA.
We further present some performance details, focussing in particular on the usage of QUDA's multigrid solver for poorly conditioned light quark monomials as well as the multi-shift solver for the non-degenerate strange and charm sector in $N_f=2+1+1$ simulations using twisted mass clover fermions, comparing the efficiency of state-of-the-art simulations on CPU and GPU machines.
We also take a look at the performance-portability question through preliminary tests of our HMC on a machine based on AMD's MI250 GPU, finding good performance after a very minor additional porting effort.
Finally, we conclude that we should be able to achieve GPU utilisation factors acceptable for the current generation of (pre-)exascale supercomputers with subtantial efficiency improvements and real time speedups compared to just running on CPUs.
At the same time, we find that future challenges will require different approaches and, most importantly, a very significant investment of personnel for software development.
}

\FullConference{The 39th International Symposium on Lattice Field Theory (Lattice2022),\\
  8-13 August, 2022 \\
  Bonn, Germany 
}

\begin{document}
\maketitle

\section{Introduction}
\vspace{-0.3cm}
Simulations of lattice QCD by the Extended Twisted Mass Collaboration (ETMC) have evolved~\cite{ETM:2015ned,Alexandrou:2018egz,ExtendedTwistedMass:2021qui} over the past decade to a point where ensembles of gauge configurations directly at the physical point can reliably be simulated at multiple lattice spacings and in large volumes~\cite{Finkenrath:2022eon} using the implementation of the Hybrid Monte Carlo (HMC) algorithm in the tmLQCD software suite~\cite{Jansen:2009xp,Abdel-Rehim:2013wba,Deuzeman:2013xaa,tmLQCD-github}, which features hybrid OpenMP/MPI parallelisation, macro-based hardware optimisations for several architectures, nested Omelyan-Mryglod-Folk~\cite{OMELYAN2003272} and force-gradient~\cite{Clark:2011ir} (FG) integrators, mass-preconditioning as well as a rational approximation for odd numbers of Wilson (clover) fermions or the non-degenerate strange and charm sector of the $N_f=2+1+1$ Wilson twisted mass (clover) action.

Moving away from preprocessor macros since about 2015, tmLQCD has been extended with interfaces to the mixed-precision CG (MP-CG) and multishift CG (MS-CG) implementations in QPhiX~\cite{joo2013xeonphi,qphixsc14,joo2016optimizing,Schrock:2015gik,QPhiX,QPhiX-github} targeting SIMD architectures and DD-$\alpha$AMG~\cite{Frommer:2013fsa,Alexandrou:2016izb,Alexandrou:2018wiv} for an efficient multigrid-preconditioned (MG-PC) solver for the most poorly conditioned monomials in our molecular dynamics (MD) Hamiltonian.
In turn, we have extended these libraries to support our Dirac operators and have shown that MG very effectively reduces the cost of simulations close to or at the physical point~\cite{Bacchio:2017pcp}.
In addition, tmLQCD has been used as glue code by various ETMC contraction codes, providing a simple interface and input file to use these libraries as well as QUDA~\cite{Clark:2009wm,Babich:2011np} and its highly efficient~\cite{Clark:2016rdz} MG-PC solver for the Wilson (clover) (twisted mass) operator.

\begin{figure}[b]
  \centering
  \includegraphics[height=52mm,clip,trim = 46mm 0 0 0]{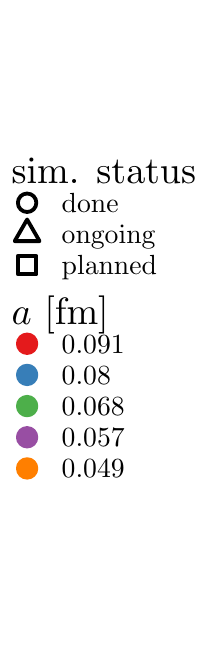}
  \includegraphics[height=52mm]{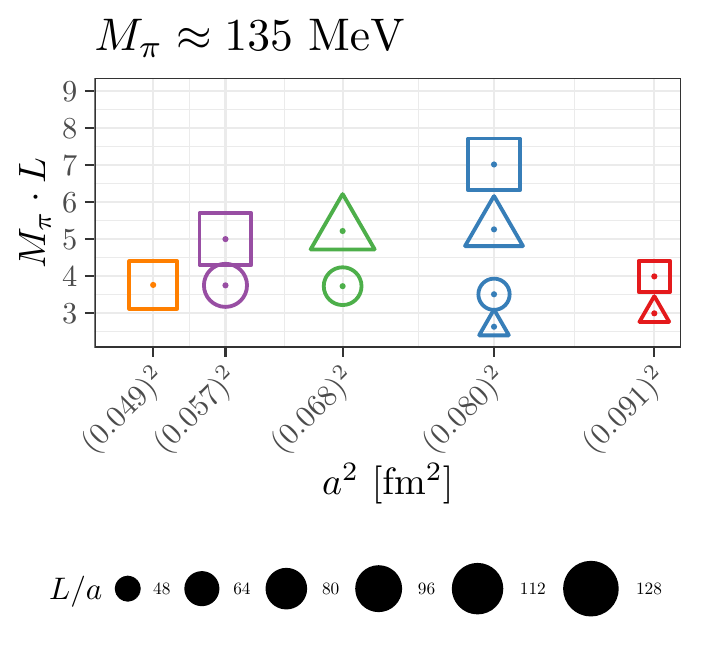}
  \includegraphics[height=52mm]{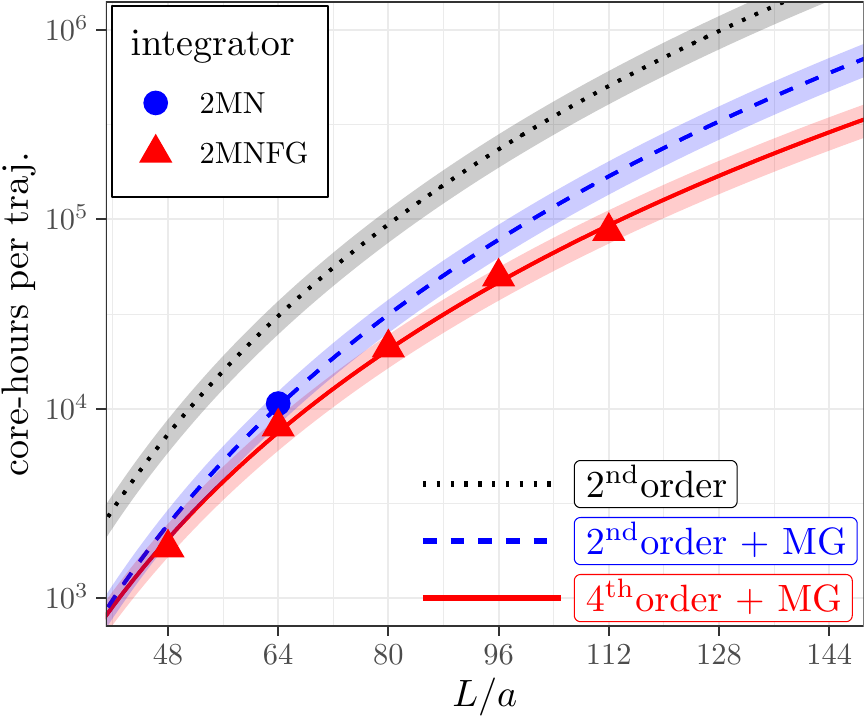}
  \caption{\it \textbf{Left:} Overview of the completed (circles), ongoing (triangles) and planned (squares) ETMC ensemble generation at the physical point. The size of the symbol indicates the lattice size. \textbf{Right:} Cost per trajectory (at the physical pion mass) in units of core hours as a function of the lattice extent in lattice units. The points stem from production simulations on different machines at roughly constant acceptance rate while the lines represent fits of the cost models given in \cref{eq:c2mncg,eq:c2mnmg,eq:c4mg}, respectively. See text body for more details. \label{fig:ensembles_phys_point}}
\end{figure}

At the precision frontier, 
observables require future ensembles with even larger volumes at many lattice spacings and with high statistics to ensure adequate control over statistical and systematic uncertainties.
Completed, ongoing and planned ETMC ensemble generation efforts in this direction directly at the physical point are shown in the left panel of \cref{fig:ensembles_phys_point}.
The right panel of the same figure gives the approximate cost per trajectory (in core hours) of generating ensembles at the physical pion mass using $N_f=2+1+1$ flavours of Wilson twised mass clover fermions as a function of the spatial lattice extent.
The computational costs for these new simulations are such that our current strategy of generating ensembles on CPU machines while employing GPU machines for physics observables is no longer tenable.
The figure relies on \emph{actual} data stemming from different machines with different generations of CPUs and simulations at different lattice spacings with comparable acceptance rates.
This data is hence subject to variations in CPU core efficiency, the number of MG iterations as well as the number integration steps required at coarser lattice spacings relative to simulations at finer ones. 
Mindful of these caveats, it is still possible to fit approximate models to the data with the black dotted, blue dashed and solid red curves given by functions of the forms:

{
\vspace{-0.6cm}
\small
\begin{align}
  C_{\textrm{2MN,CG}} \approx & c_2 \cdot {(V/a^4)}^{5/4} \label{eq:c2mncg} \\
  C_{\textrm{2MN,MG}} \approx & c_2' \cdot {(V/a^4)}^{5/4} \label{eq:c2mnmg} \\
  C_{\textrm{2MNFG,MG}} \approx & c_4 \cdot {(V/a^4)}^{9/8} \label{eq:c4mg} \,,
\end{align}
}

\vspace{-0.2cm}
corresponding to three types of setups: MP-CG only with a nested second order integration scheme; the same scheme but employing DD-$\alpha$AMG for the most poorly conditioned monomials; that same solver setup but employing a FG scheme instead (and appropriate mass-preconditioning).
Even with the best setup, a $128^3\cdot256$ ensemble costs in excess of $2 \cdot 10^5$ core-hours per trajectory and, when run at a reasonable parallel efficiency, already a $96^3\cdot192$ ensemble entails a real time per trajectory of around six hours.
Thus, not only would all currently planned runs require $\mathcal{O}(10^9)$ core-hours, too much to apply for on available CPU machines, they would also take too much real time to perform without resorting to generating multiple independent Markov chains, complicating subsequent analysis. 
With the actual or imminent availability of (pre-)exascale machines based on accelerated architectures by various vendors in mind, porting our HMC implementation to these is thus not only necessary but also timely.

\section{Even-odd and mass-preconditioned Dirac operators and framework choice}
\vspace{-0.3cm}
When used as a driver for inversions of the full Dirac operator from within contraction codes, tmLQCD's QUDA interface delegates even-odd preconditioning completely to QUDA by simply setting \texttt{solve\_type} and \texttt{matpc\_type} of \texttt{QudaInvertParam} correctly for the given solver\footnote{It should be noted that QUDA's parameter set is very large and that (in most cases) a gamma basis change is required.}.
In the HMC, however, the even-odd (and mass) preconditioning in the action must match that used in QUDA.
To be specific, our degenerate determinant is written in terms of operators $\Qpm$,
\begin{align}
  \Qpm = & \gamma_5 (M_{\text{clov}} \pm i \mu_\ell \gamma_5) \;\; \text{s.t.} \;\; {(\Qp)}^\dagger = \Qm \\
  \det( \Qp \Qm ) = & \det( Q^2 + \mu_\ell^2 )\,,
\end{align}
where $M_{\text{clov}}$ is the Wilson-clover Dirac operator and $\mu_\ell$ is the light twisted quark mass.
We employ asymmetric even-odd preconditioning\footnote{\texttt{ODD\_ODD\_ASYMMETRIC} in QUDA-parlance} and thus require an implementation of
\begin{equation}
  \hQpm = \gamma_5 \left[ ( \Moo \pm i \mu_\ell \gamma_5 ) - \Moe {(\Mee \pm i \mu_\ell \gamma_5 )}^{-1} \Meo \right] \label{eq:hQpm}
\end{equation}
and for mass-preconditioning, we further require
\begin{equation}
  \hWpm(\rho) = \hQpm \pm i \rho \;\; \text{s.t}. \;\; \hWp(\rho) \hWm(\rho) = \hQp \hQm + 2\mu_\ell \rho + \rho^2 \label{eq:hWpm}
\end{equation}
where $\rho$ is the preconditioning mass, leaving the inverse of the clover term $\rho$-independent.
In the heavy sector we need a two-flavour operator in which the diagonal depends on both $\bar{\mu}$ and $\bar{\epsilon}$:
\begin{equation}
  \hQh = \gamma_5 \left[ ( \Moo + i \bar{\mu} \gamma_5 \tau^3 - \bar{\epsilon} \tau^1 ) - \Moe {(\Mee + i \bar{\mu} \gamma_5 \tau^3 -     \bar{\epsilon} \tau^1 )}^{-1} \Meo \right] \,. \label{eq:hQh}
\end{equation}

When we evaluated different frameworks for implementing our HMC for GPU machines, we considered essentially three factors: availability of a proven-efficient GPU implementation of a MG-PC solver; ease of implementation of \cref{eq:hQpm,eq:hWpm,eq:hQh} in the given framework and the availability of a flexible input format for setting up the HMC and RHMC components required for our simulations.
Although doing so would have contributed towards solving the wider performance-portability challenge of various ETMC codebases, time limitations ruled out extending Grid~\cite{Boyle:2016lbp,Richtmann:2019eyj,Richtmann:2022fwb} or the Chroma+QDP-JIT+QUDA tripos~\cite{Edwards:2004sx,Winter:2011np,Winter:2014npa,chromaqdpjit}.

The efforts of the QUDA development team towards a device target abstraction and the resulting development or availability of backends for CUDA, HIP, SYCL and OpenMP, clearly favoured a compromise of proceeding with tmLQCD + QUDA for the time being without solving the more general issue.
As first steps, we only had to implement: more fine-grained tracking of the state of the host and device gauge fields (and related objects such as the clover term and MG setup); \cref{eq:hWpm,eq:hQh} as well as an extension of tmLQCD's QUDA interface to support the kinds of solves required within the HMC\@.
We further enabled the gauge derivative to be offloaded and helped to modify QUDA's gradient flow interface such that the values of pertinent flowed observables can be returned to the host application (if desired) in the addition to the flowed gauge field.

\section{Laying the groundwork}
\vspace{-0.3cm}
Before the main implementation phase, we extended the mechanism by which tmLQCD tracks the state of its gauge field (and the various copies thereof, as well as their halos).
To this end, we introduced the concept of a \texttt{tm\_GaugeState\_t} struct at program scope which contains a \texttt{gauge\_id} member, corresponding simply to the position along a trajectory in the HMC, expressed as a real number\footnote{or the gauge configuration number when tmLQCD is used as a solver driver from within contraction codes}.
Whenever the gauge field is updated, the function \texttt{update\_tm\_gauge\_id} is called and functionality exists which cascades to the state of the clover field and its inverse.
A similar set of states and functions were introduced to the QUDA interface to track the state of the device gauge and clover fields as well as the MG setup, such that these can be refreshed as necessary.
At the cost of efficiency, this allowed us to keep the representative gauge and conjugate momentum fields on the host and to add functionality step by step, allowing us to test our developments while being able to mix QUDA functions with our own CPU functions in different places without having to worry about breaking any synchronisation assumptions.

As a further preparatory step, we implemented a simple profiling mechanism to quantify relative costs associated with different parts of our MD Hamiltonian combined with hierarchical information about the function call tree.
While it can be argued that the latter can be obtained using readily available profiling tools, our wish to profile full scale simulations with $\mathcal{O}(10)$ monomials employing MG-PC, MP-CG and MS-CG solvers on GPU machines makes this ineffective without explicit (profiler-specific) annotations.
Since functions are often called from multiple monomials and sometimes even at different levels of the call tree (depending on context), it may be difficult to differentiate between physically and algorithmically reasonable hot spots and those caused by incorrect parameter / algorithm choices or implementation / interfacing mistakes. 
The new stack-based profiler is used through just two functions: \texttt{tm\_stopwatch\_push} and \texttt{tm\_stopwatch\_pop}.
The former starts a timer at a given level of the call tree and annotates it with (caller-defined) context information (similar to UNIX paths), while the latter stops the timer of the current level and prints the measured time to \texttt{stdout} with the aforementioned context information.

\begin{SCfigure}
  \centering
  \includegraphics[height=60mm]{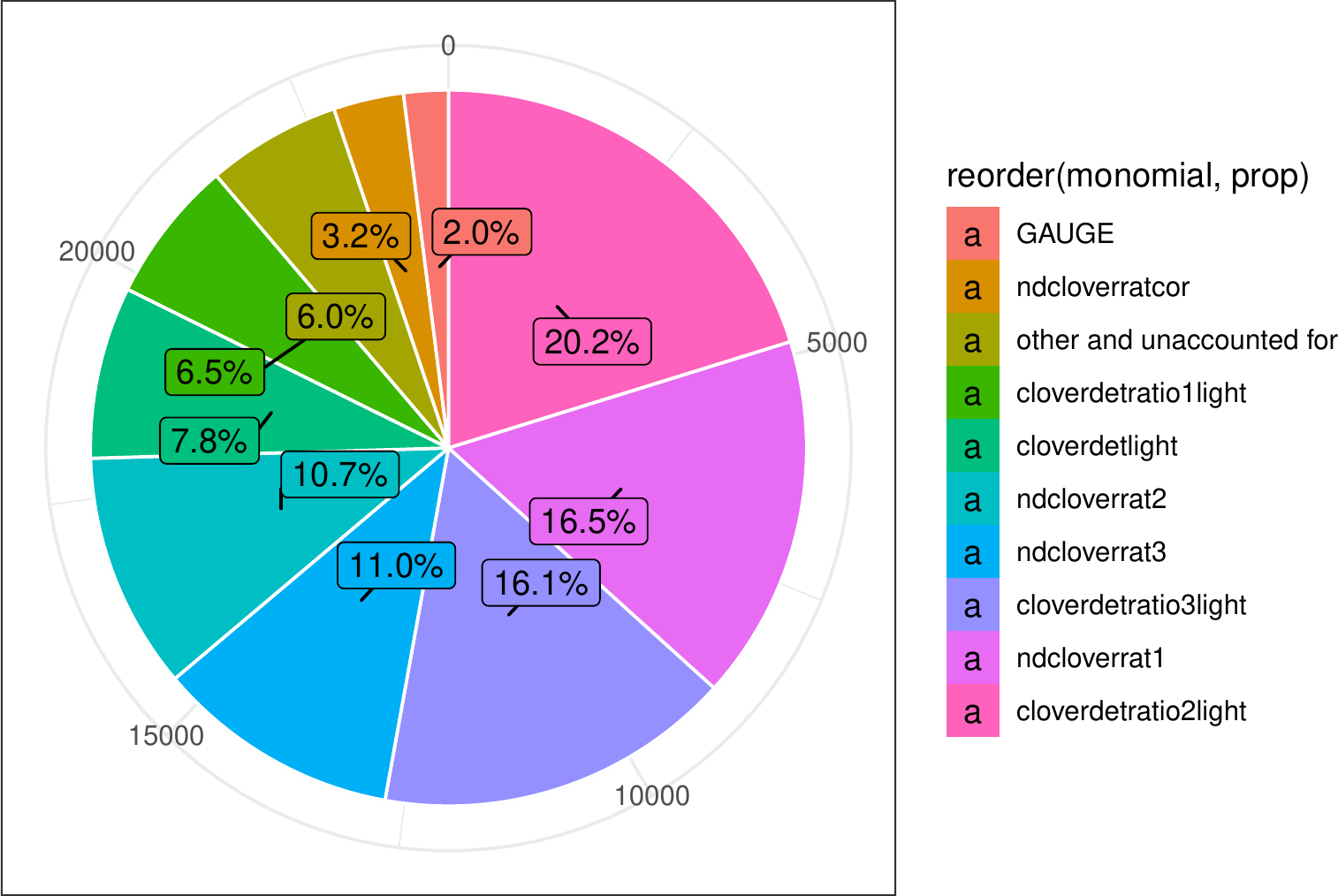}
  \caption{\it Profile of tmLQCD's HMC running multiple trajectories of a $N_f=2+1+1, V/a^4 = 64^3\cdot128$ ensemble at the physical point on 16 Marconi-100 nodes (4 NVIDIA V100 GPUs per node, 2 $\times$ 16-core IBM POWER9 AC922 CPUs per node). The absolute time in seconds spent in each monomial is shown as well as what fraction of the total this corresponds to. \label{fig:hotspots_by_monomial}}
\end{SCfigure}

As a specific example, the function \texttt{cloverdet\_derivative}, which calculates the force due to a Wilson clover (twisted mass) fermion determinant, is bracketed by such calls:
\begin{lstlisting}
tm_stopwatch_push(&g_timers, __func__, mnl->name);
[...]
tm_stopwatch_pop(&g_timers, 0, 1, "");
\end{lstlisting}
and the same applies to almost all functions further down the call tree, giving output like:
\begin{lstlisting}
# TM_QUDA: Time for invert_eo_degenerate_quda 2.937360e-01 s level: 3 proc_id: 0 /HMC/cloverdetlight:clover[...] 
# : Time for gamma5 5.561520e-04 s level: 3 proc_id: 0 /HMC/cloverdetlight:cloverdet_derivative/solve_degen[...]
[...]
# : Time for sw_all 4.972867e-01 s level: 2 proc_id: 0 /HMC/cloverdetlight:cloverdet_derivative
# : Time for cloverdet_derivative 1.252163e+00 s level: 1 proc_id: 0 /HMC/cloverdetlight:cloverdet_derivative
\end{lstlisting}
An R script is provided in \texttt{profiling/hmc\_mk2} which generates a PDF report from such a log file (even if it is incomplete) visualising and tabulating the distribution of the time spent in various parts, as shown exemplarily in \cref{fig:hotspots_by_monomial}, where the different labels correspond to different monomials in our MD Hamiltonian (going into the details is beyond the scope of the present proceedings).

\section{Implementation details and performance}
\vspace{-0.3cm}
\subsection{Multigrid solver in the light sector}
\begin{SCfigure}
  \includegraphics[width=6.8cm]{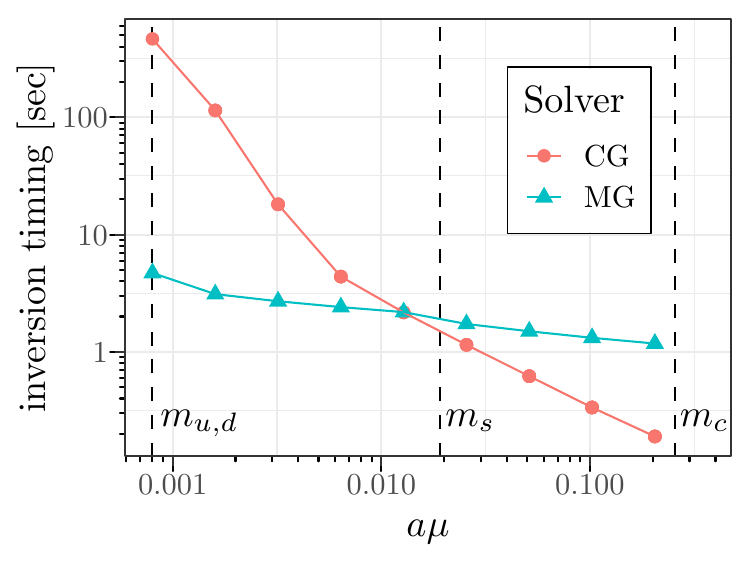}
  \caption{\it QUDA-MG and mixed-precision CG compared as a function of the twisted quark mass. The timings refer to a single inversion in the case of CG and two inversions in the case of MG (including some unavoidable overheads), as required in the calculation of the fermion derivative. The dashed vertical lines indicate the physical light and approximate strange and charm quark masses. \label{fig:mg_vs_cg}}
\end{SCfigure}

\vspace{-0.1cm}
The importance of using a MG-PC solver in simulations close to or at the physical point can hardly be overstated.
While one could worry about the (minor) setup overhead at the start of a run and due to setup refreshes or updates of the coarse-grid operators, the algorithmic superiority more than makes up for these.
As shown in \cref{fig:mg_vs_cg}, in computing the fermionic derivative, QUDA's MG is faster than its MP-CG by about a factor of 100 at the physical light quark mass.
In practice, we optimize costs by employing a mixture of solvers for different monomials in the light sector, usually MG for the two most poorly conditioned determinant ratios and MP-CG everywhere else, leading to an overall whole-application speedup of about a factor of four compared to using MP-CG alone. 

Depending on the type of mass preconditioning employed, using QUDA-MG in the HMC requires some care.
To be specific, when $\rho$ in \cref{eq:hWpm} is 20 to 30 times larger than the target quark mass, the setup fails to precondition the fine grid operator well, resulting in hundreds of outer GCR iterations.
In determinant ratios with large $\rho$ in the denominator, we counter this by employing MP-CG in the heatbath step while still using MG in the derivative and acceptance step. 
With more effort, as was done for DD-$\alpha$AMG~\cite{Alexandrou:2016izb}, QUDA-MG could be modified to more effectively deal with \cref{eq:hWpm}. 
Another inefficiency which can perhaps be resolved in the future, is that coarsening currently only supports the symmetrically even-odd-preconditioned operator.
We simply ignore this and set:
\begin{itemize}
  \item \texttt{matpc\_type = QUDA\_MATPC\_ODD\_ODD} in \texttt{QudaMultigridParam.invert\_param}
  \item \texttt{matpc\_type = QUDA\_MATPC\_ODD\_ODD\_ASYMMETRIC} in the outer solver, 
\end{itemize}
likely resulting in slightly higher iteration counts than with a more consistent setup.

\subsection{Mixed-precision multi-shift solver in the heavy sector}
\begin{figure}
  \begin{minipage}{0.4\textwidth}
    \includegraphics[height=50mm]{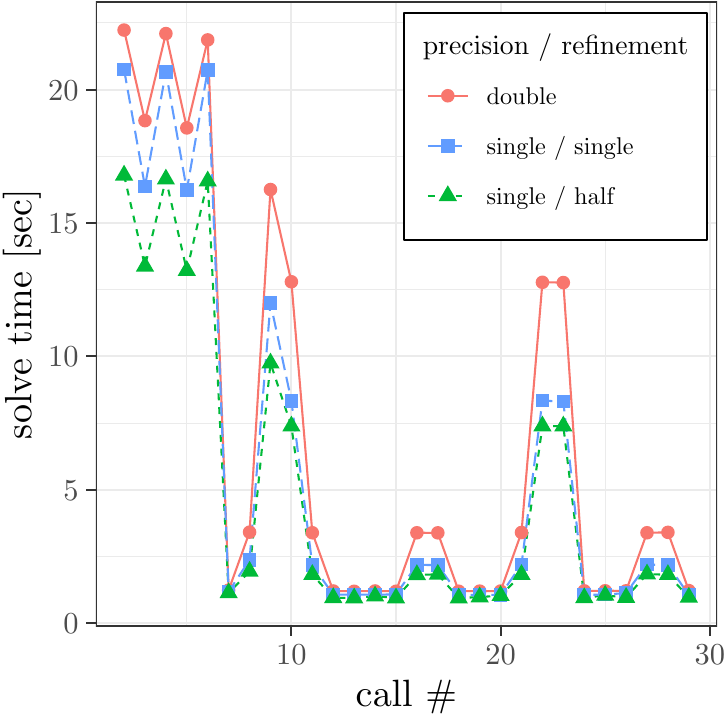}
  \end{minipage}
  \begin{minipage}{0.59\textwidth}
    {\footnotesize
    CPU: 3072 Intel Xeon Platinum 8168 cores (64 Juwels nodes) \\
    GPU: 32 A100 + 384 EPYC Rome 7402 cores (8 Juwels Booster nodes) \\[2ex]
    }
    {\footnotesize
    \begin{tabular}{lll}                                                                                                             
      Machine / Algorithm & HB & ACC \\
      \hline \hline \\
      (CPU) QPhiX multi-shift CG & 810 s & 550 s \\
      (CPU) DD-$\alpha$AMG accelerated multi-shift CG & 590 s & 400 s \\
      (GPU) QUDA mshift CG (double) & 145 s & 93 s \\  
      (GPU) QUDA mshift CG (single / single) & 127 s & 79 s \\  
      (GPU) QUDA mshift CG (single / half) & 103 s & 66 s 
    \end{tabular}
    }
  \end{minipage}

  \caption{\it \textbf{Left:} Timings of MS-CG calls along a trajectory ($64^3 \cdot 128$ lattice) running on 32 A100 GPUs (8 Juwels Booster nodes) using full double precision (red circles), single precision MS-CG with double-single shift-by-shift refinement (blue squares) or single precision MS-CG with double-half shift-by-shift refinement (green triangles). \textbf{Right:} Comparison of timings of the rational approximation correction term in the same simulation between running on 3072 CPU cores (64 Juwels nodes) and 32 A100 GPUs. \label{fig:quda_multishift_speedup}}
\end{figure}

\vspace{-0.1cm}
As can be seen in \cref{fig:hotspots_by_monomial}, a significant fraction of total runtime is spent in monomials simulating partial fractions of the rational approximation of the non-degenerate $1+1$ sector of our action (\texttt{ndcloverratN} and \texttt{ndcloverratcor} in the figure).
In simulations on CPU machines, the calculation of the correction term for the rational approximation~\cite{PhysRevLett.98.051601,luescher_computational_strategies} contributes significantly to the total cost.
This is shown in the table of \cref{fig:quda_multishift_speedup}: using only QPhiX MS-CG takes about 1360 seconds and dropts to around 990 seconds employing DD-$\alpha$AMG refinement~\cite{Alexandrou:2018wiv}.
On 32 NVIDIA A100 GPUs instead, this takes only 220 seconds using double precision MS-CG and improves down to around 170 seconds when single precision MS-CG with double-half shift-by-shift refinement is used.
Similar improvements are seen in all calls of MS-CG, as shown in the left panel of \cref{fig:quda_multishift_speedup}.

\subsection{Efficiency}
\begin{figure}
  \includegraphics[width=0.49\textwidth]{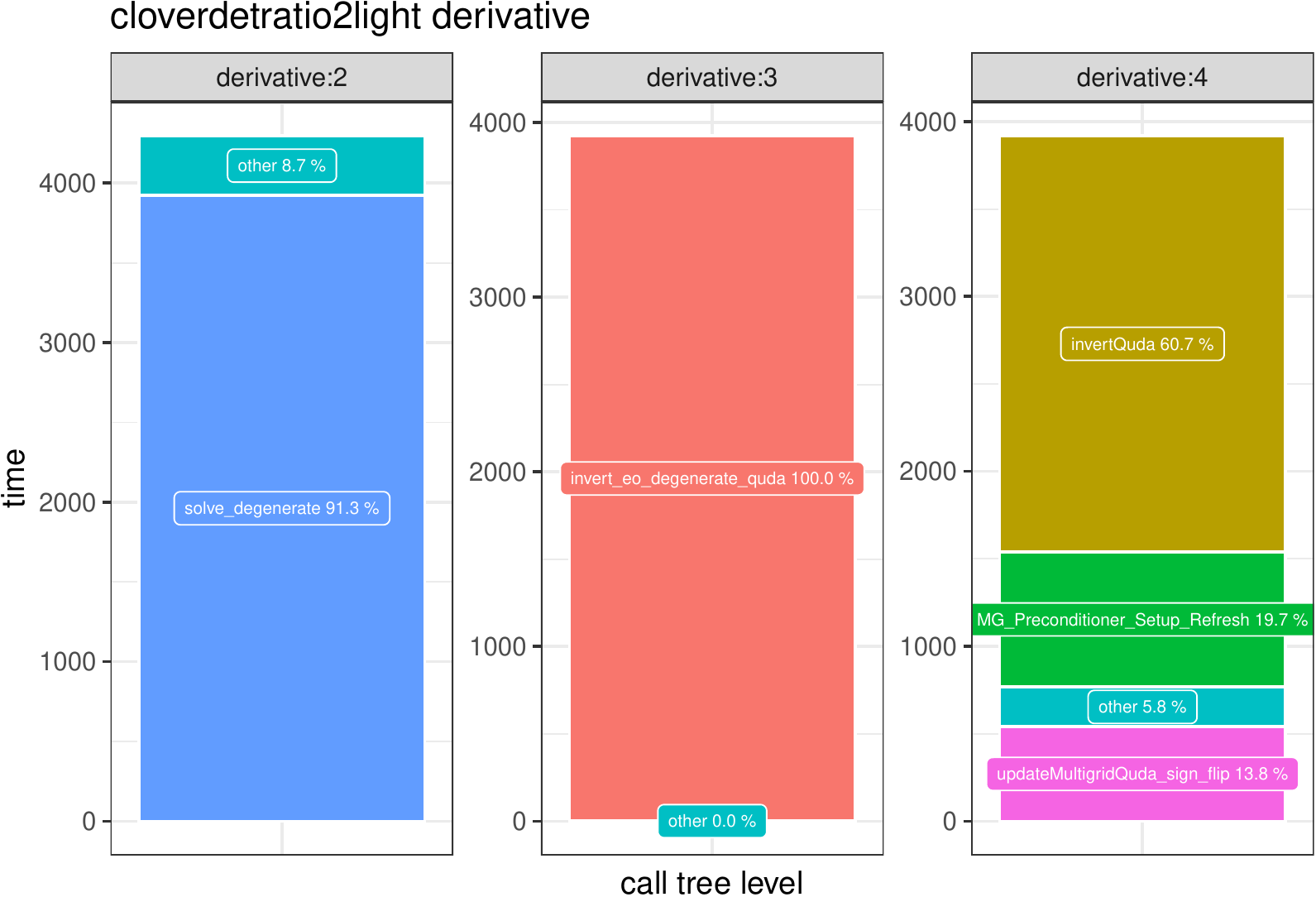}\hfill
  \includegraphics[width=0.49\textwidth]{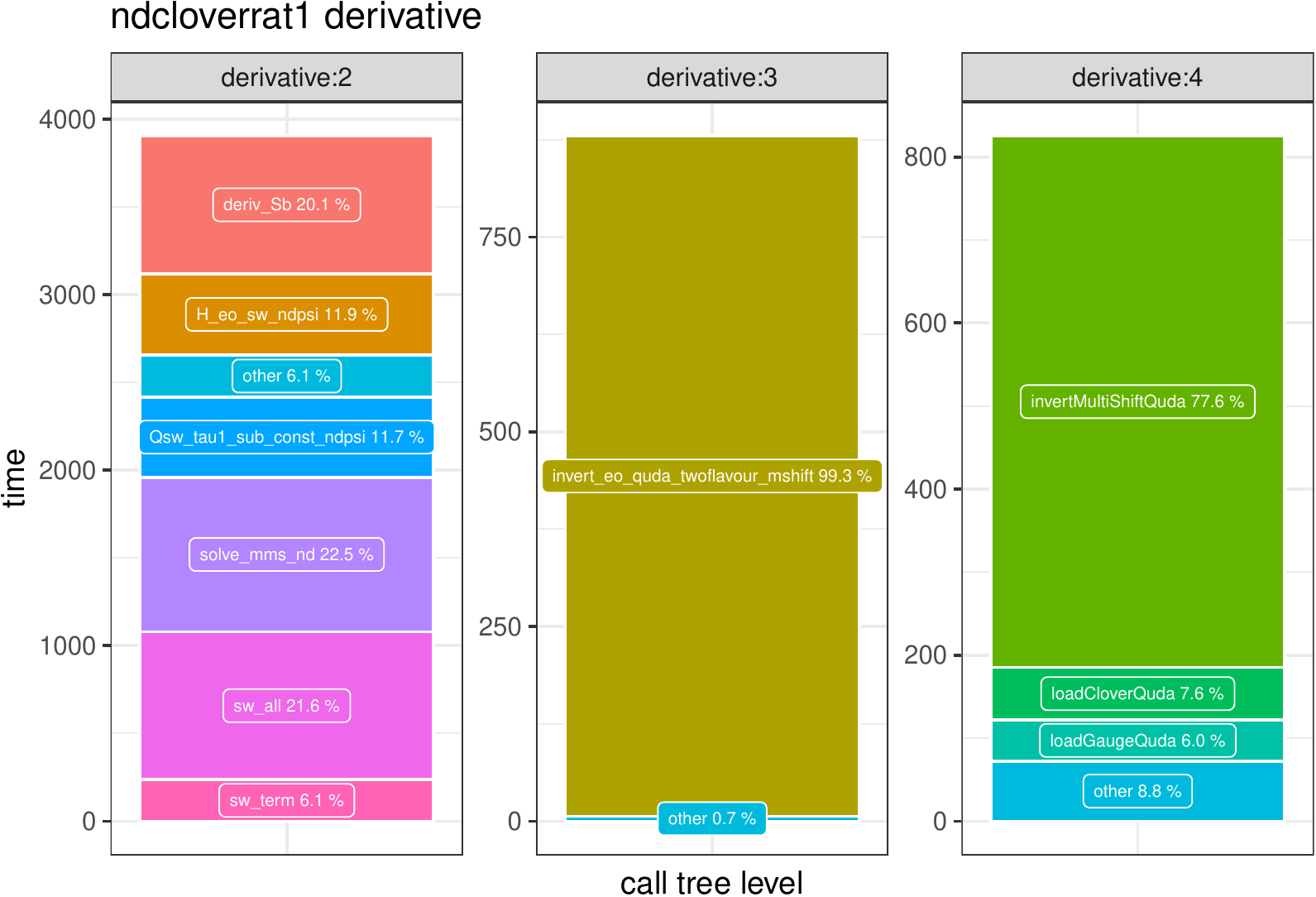}
  \caption{\it \textbf{Left:} Hierarchical profile of the force calculation of a monomial (see also \cref{fig:hotspots_by_monomial}) dominated by the \texttt{solve\_degenerate} function running mostly on the GPU (specifically \texttt{invertQuda} and unavoidable MG overheads). \textbf{Right:} Profile of a monomial derivative where only about 20\% is offloaded to the GPU (\texttt{solve\_mms\_nd}). \label{fig:hierarchical_profile_monomial}}
\end{figure}

\vspace{-0.1cm}
In our profiles, we can identify monomials which are very much GPU-dominated as shown in the left panel of \cref{fig:hierarchical_profile_monomial}, where almost all of the time is spent in the solver driver (\texttt{solve\_degenerate}), the corresponding QUDA function (\texttt{invertQuda}) or unavoidable overheads related to the MG solver.
On the other hand, as shown in the right panel of the same figure, we still have monomials with very poor offloading fractions.
In the force calculation for the rational approximation, outer products must be computed on multiple two-flavour spinors and for the largest shifts, the solver requires only very few iterations.
Since we have not yet extended QUDA to support exactly what we require, these outer products on the CPU thus dominate total cost in this case.

\begin{figure}
  \begin{minipage}{0.5\textwidth}
    \centering
    \includegraphics[height=50mm]{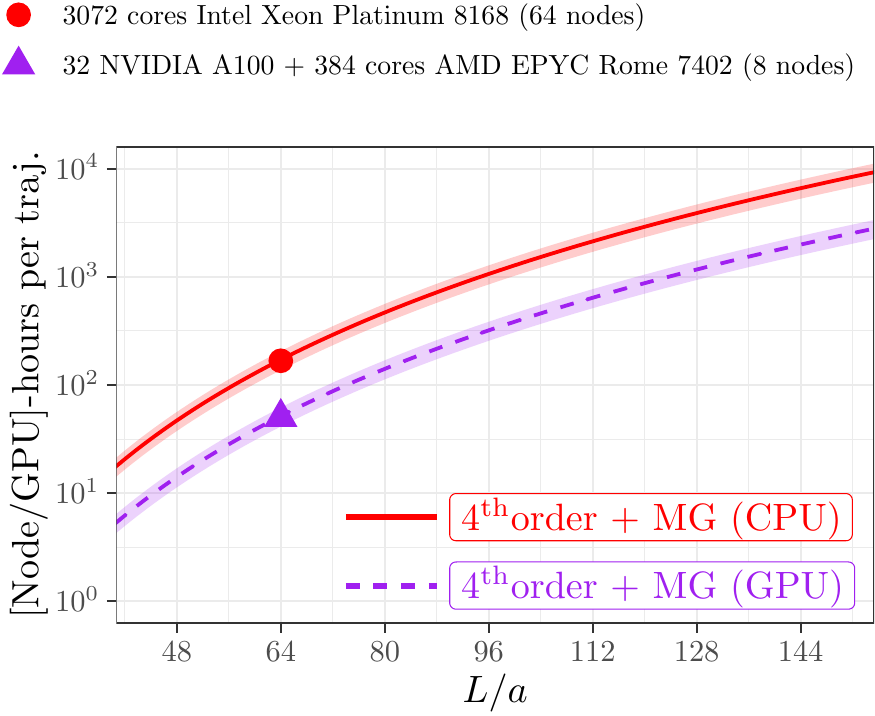}
  \end{minipage}
  \begin{minipage}{0.50\textwidth}
    \footnotesize
    \begin{tabular}{cccc}
      machine                  & real  & CPU node-hrs / &    kWh      \\                                                        
                               & time  & GPU-hrs          &             \\
      \hline \\
      64 nodes         &  $2.61$ h  & 167                        &    $\sim 84$      \\  
      (Juwels)         &            &                            &             \\
                       &            &                            &             \\
      32 GPUs          &  $1.58$ h  & 50.6                       &    $\sim 24$    \\    
      (Juwels Booster) &            &                            &    
    \end{tabular}
  \end{minipage}
  \caption{\it \textbf{Left:} Cost per trajectory for a $64^3\cdot128$ simulation at the physical point (in node-hours for the CPU machine, GPU-hours for the GPU machine). \textbf{Right:} Real time and cost per trajectory on 64 Juwels nodes and 8 Juwels Booster nodes, respectively. Estimate of the respective power usage per trajectory. \label{fig:ndhrs_per_traj}}
\end{figure}

Even with these inefficiencies we reach overall GPU utilisation fractions between 30 and 70\% depending on the machine, the size of the machine partition and the simulation parameters (larger partitions, lighter quark mass $\rightarrow$ higher utilisation).
In order to \emph{attempt} an apples-to-apples comparison, \cref{fig:ndhrs_per_traj} shows the cost per trajectory in node-hours on 3072 Intel Xeon Platinum 8168 cores (64 Juwels nodes) compared to the cost in GPU-hours on 32 NVIDIA A100 (8 Juwels Booster nodes) for a $64^3 \cdot 128$ ensemble at the physical point.
Since a Juwels node uses about a factor of four less power than a Juwels Booster node, we can estimate that in addition to reducing real time by a factor of around 1.7, energy efficiency has improved by about a factor of three, with further improvement expected when the remaining parts of the fermionic force are offloaded.

\newpage
\subsection{Strong-scaling and performance-portability}
\vspace{-0.1cm}
\begin{wrapfigure}{r}{0.40\textwidth}
  \centering
  \includegraphics[width=0.39\textwidth]{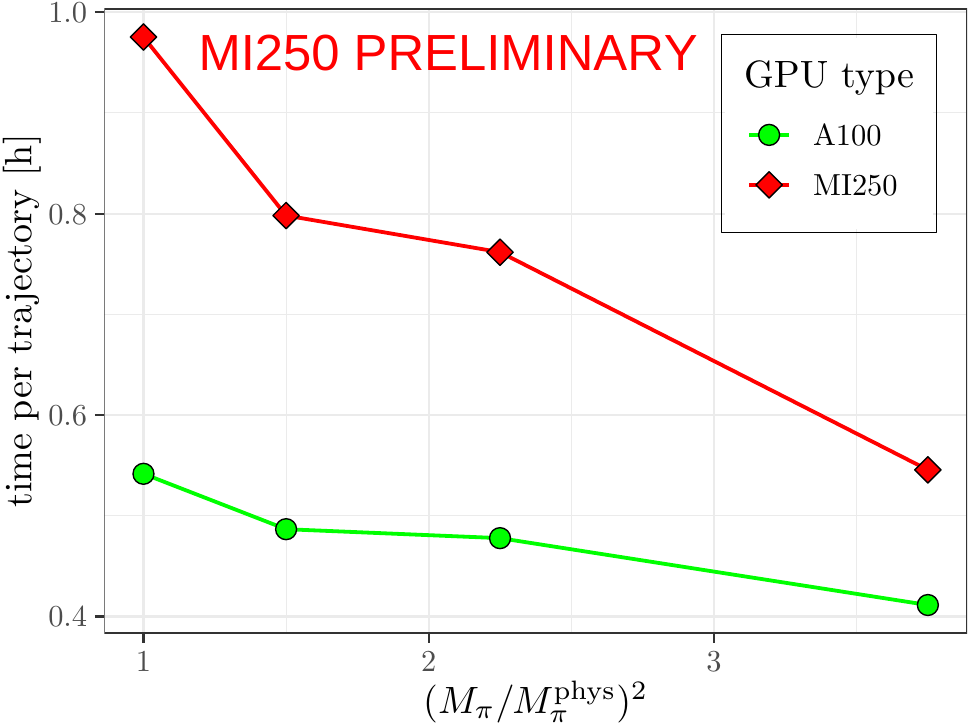}
  \caption{\it Time per trajectory for well-thermalised $N_f = 2+1+1, V/a^4 = 32^3\cdot64$ simulations from about $M_\pi \sim 250\,\text{MeV}$ down to the physical point (using the same number of integration steps for all pion masses) on a single Juwels Booster node with 4 A100 GPUs (green circles) and a single node of a test system at JSC with 4 MI250 GPUs. \label{fig:hip_vs_cuda}}
\end{wrapfigure}
\begin{SCfigure}
  \includegraphics[width=0.33\textwidth,page=1]{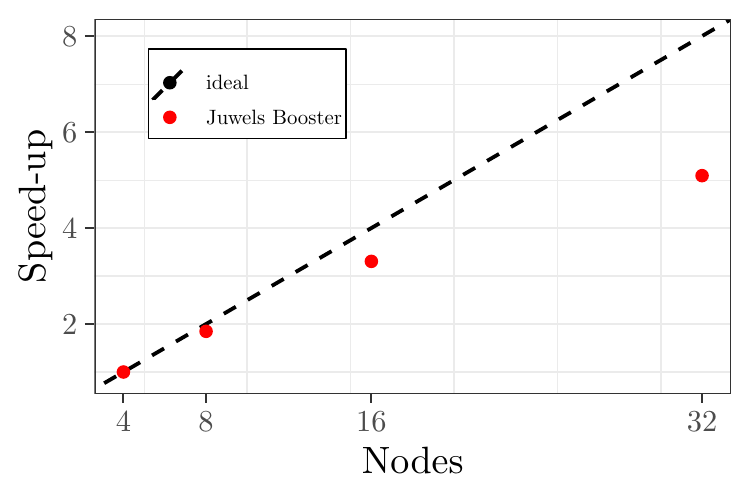}
  \includegraphics[width=0.33\textwidth,page=2]{plots/Juwels_Booster_HMC_Scaling}
  \caption{\it \textbf{Left:} Strong-scaling of a $64^3\cdot128$ simulation at the physical point from 4 to 32 Juwels Booster nodes. \textbf{Right:} Strong-scaling of a $112^3\cdot224$ simulation at the physical point from 28 to 112 Juwels Booster nodes. \label{fig:strong_scaling}}
\end{SCfigure}

In \cref{fig:strong_scaling} we show full-application behaviour for a $64^3\cdot128$ simulation scaled from 4 to 32 Juwels Booster nodes and a $112^3\cdot224$ simulation scaled from 28 to 112 nodes.
While strong scaling is quite good, it will worsen once the remaining parts of the fermionic force are offloaded as it will become more dominated by the scalability of the MG algorithm.

Finally, to demonstrate the performance-portability of our approach, in \cref{fig:hip_vs_cuda} we show the time per trajectory of thermalised $32^3\cdot64$ simulations running at different pion masses on a single node of Juwels Booster (4 NVIDIA A100 GPUs) and a single node of a \emph{preliminary} test system at JSC (4 AMD MI250 GPUs).
It should be noted that the environment (ROCm version, drivers, software stack) on the AMD-based test system was not at all final and that none of the algorithmic parameters, in particular the MG setup, were retuned for the MI250-based machine.
Still, with just a little bit of work on the tmLQCD \emph{build system}, we were able to run tmLQCD + QUDA unmodified with good performance on an architecture wich we had never used before.

\section{Conclusions}
We have shown in this contribution that employing QUDA through its C interface to offload the most expensive parts of the HMC is suitable for the current generation of GPU-accelerated supercomputers.
Although we still need to offload a number of parts of the fermion force, we expect to eventually be able to reach GPU utilisations between 60 and 70\% on (pre-)exascale machines for state-of-the-art simulations.
Because we are dependent on a backend being available for a particular architecutre, QUDA is likely not a general solution for our performance-portability problem, for instance to target future CPU generations.
We will also run into issues if very dense GPU configurations are coupled with weak driver CPUs to keep total power consumption under control, as we would need to offload a much larger fraction of tmLQCD's functionality, necessitating support for different memory layouts and execution spaces and thus essentially a complete rewrite.
Further improvements to scalability could be obtained through task parallelism (e.g.\ multiple monomials on the same time scale, the n-th root trick or through modular supercomputing architectures~\cite{suarez2019modular}) at the cost of substantial refactoring.
These improvements could be achieved by moving to Grid or perhaps by using Kokkos~\cite{CARTEREDWARDS20143202} as a performance-portability layer.
In addition, the interfacing layer provided by Lyncs~\cite{Bacchio:2022bjk} could allow combining different libraries (Grid and QUDA, for example) and enable both task parallelism and higher programmer productivity.
All of these challenges, however, require a significant personnel investment into software development for lattice field theory, ideally as a community effort.
\vspace{-0.3cm}
\acknowledgments
\vspace{-0.3cm}
{\footnotesize
We would like to thank the QUDA developers for their tremendous work as well as the many pleasant and productive interactions during this and previous efforts.
We thank the ETMC for the most enjoyable collaboration.
B.K. was funded by HPC\@.NRW\@. for part of this work.
S.B. and J.F. are supported by the H2020 project PRACE 6-IP (grant agreement No. 82376) and the EuroCC project (grant agreement No. 951740). 
We acknowledge support by the European Joint Doctorate program STIMULATE grant agreement No. 765048.
This work is supported by the Deutsche Forschungsgemeinschaft (DFG, German Research Foundation) and the NSFC through the funds provided to the Sino-German Collaborative Research Center CRC 110 “Symmetries and the Emergence of Structure in QCD” (DFG Project-ID 196253076 - TRR 110, NSFC Grant No. 12070131001).
The authors gratefully acknowledge the Gauss Centre for Supercomputing e.V. (www.gauss-centre.eu) for funding this project through computing time on the GCS supercomputer JUWELS Booster~\cite{JUWELS} at the J{\"u}lich Supercomputing Centre.
Some of the runs done for this work were carried out on the Bender GPU cluster at the University of Bonn and we gratefully acknowledge support by the HRZ-HPC team.
}

\bibliographystyle{h-physrev5}

\end{document}